\documentclass[10pt,draftclsnofoot,journal,twocolumn]{IEEEtran}
\usepackage{cite,url,subfigure,epsfig,graphicx,wrapfig,psfrag} 
\usepackage{amsmath}
\usepackage{multirow}
\usepackage{amssymb}
\usepackage{algorithmic,algorithm}
\usepackage{authblk}

\begin{document}

\title{
Efficient Data Exchange in Unmanned Aerial Vehicle Networks Utilizing Unsupervised Learning-Based Clustering}

\author{Hao Song, Lingjia Liu, and Ananth Balasubramanian
\thanks{H. Song is with the Next Generation and Standards (NGS) group, Intel Corporation, Santa Clara, CA 95054, USA (hao.song@intel.com).
L. Liu is with Bradley Department of Electrical and Computer Engineering, Virginia Tech, Blacksburg, VT 24060, USA. 
A. Balasubramanian is with Qualcomm Inc, San Diego, USA.
The corresponding author is L. Liu (ljliu@ieee.org).}
}

\maketitle

\begin{abstract}

An unmanned aerial vehicle (UAV) network can serve as an aerial relay to periodically receive packets from macro base stations (BSs). 
Severe packet loss may happen especially when UAVs have bad wireless connections to a BS. 
In this paper, a data exchange scheme is proposed utilizing unsupervised learning to enable efficient lost packet retrieval through reliable wireless transmissions between UAVs instead of through retransmissions of macro BSs with a longer delay and higher overhead. 
With the proposed scheme, all UAVs are assigned to multiple clusters and a UAV can only request its lost packets to UAVs in the same cluster. By this way, UAVs in different clusters could carry out their lost packets retrieval processes simultaneously to expedite data exchange. The agglomerative hierarchical clustering, a type of unsupervised learning, is used to conduct clustering, guaranteeing that UAVs clustered together could supply and supplement each other's lost packets. To further enhance data exchange efficiency, a data exchange mechanism is designed, where the priority of UAVs performing data exchange is determined by the number of their lost packets or the number of requested packets that they can provide. The introduced data exchange mechanism can make each request-reply process maximally beneficial to other UAVs' lost packet retrieval in the same cluster. A new random backoff procedure based on the carrier sense multiple access with collision avoidance (CSMA/CA) is designed to support the data exchange mechanism. Finally, simulation studies are conducted to demonstrate the effectiveness and superiority of our proposed data exchange scheme.

\end{abstract}

\section{Introduction}

\IEEEPARstart{U}{nmanned} aerial vehicle (UAV) networks have wide-ranging applications in wireless communications of both military and commercial field because of their significant advantages. High mobility, high flexibility, and low complexity make UAV networks capable of supporting wireless access services when wireless communications infrastructures, like base stations (BSs), are not available or cannot work \cite{UAVSurvery1, UAVSurvery2, UAVSurvery3}. For example, UAV networks are able to provide emergency communications in disaster areas where local infrastructures have already got crippled \cite{UAVEmergency, UAVEmergency1}. Moreover, compared to deploying costly infrastructures, UAV networks are a cost-effective alternative to provide wireless coverage in some special scenarios. For example, UAV networks could be used in sensor networks, flying to sensor nodes and collecting data from them \cite{UAVSensor}. UAVs are also useful to provide efficient operations in some mechanical motions and routine missions, such as traffic monitoring \cite{UAVRoutineMission1} and agricultural surveillance \cite{UAVRoutineMission2}. Besides, UAVs could enhance communication quality and system capacity by being deployed in communication hotspots and areas with weak signal strength \cite{UAVCellular}.

In this paper, we focus on studying the application of utilizing UAV networks as small aerial BSs, where UAV networks serve as relays, forwarding data from macro BSs to users \cite{UAVSurvery1, UAVSurvery2, UAVCellular}. In such an application, UAVs need to periodically receive packets from macro BSs. Additionally, it is likely for UAV networks to be deployed far away from macro BSs to cover areas where macro BSs cannot provide effective coverage. Wireless transmissions between UAVs and macro BSs in a long-distance propagation cause weak received signal strength and high lost data rate. Traditionally, lost packets are retrieved through retransmissions of macro BSs \cite{Retransmission1, Retransmission2}. After a UAV receives a packet from a macro BS, it will feed acknowledgement (ACK) back to the macro BS. If the macro BS found that there still are UAVs not receiving transmitted packets according to ACK feedback record, the macro BS will keep retransmitting the packets until all UAVs successfully receive them \cite{Retransmission3}. Apparently, in UAV networks, lost packets retrieved through retransmissions of macro BSs may cause a severe delay because of long-distance transmissions and a large amount of UAVs. To be specific, when UAV networks work in a bad wireless environment with high lost data rate, for example located far away from macro BSs, multiple retransmissions may have to be conducted before all UAV successfully receive a packet with a large delay. Moveover, the delay issue may be deteriorated with a large amount of UAVs, as more retransmissions may be encountered with more UAVs. Besides a severe delay, enormous retransmissions may result in serious overhead and power consumption for both UAVs and macro BSs. 
Essentially, enabling feedback in broadcast communications with large number of users becomes inefficient.
This is because UAVs are required to feed back an ACK for each received packet, while the BS has to record and process enormous ACKs from all UAVs.

In a UAV network working as a small aerial BS, UAVs normally stay together, working as a swarm to handle complicated missions \cite{SwarmUAV1, SwarmUAV2, SwarmUAV3}. Thus, wireless transmissions between UAVs in a short distance would be more reliable. It would also be more friendly to delay, overhead, and power if the lost packet retrieval can be conducted by data exchange between UAVs rather than by retransmissions of macro BSs. In this paper, an data exchange scheme leveraging unsupervised learning is proposed in UAV networks, enabling intelligent and efficient data exchange between UAVs. In the proposed data exchange scheme, a UAV network is partitioned into multiple clusters using the agglomerative hierarchical clustering, a type of unsupervised learning \cite{HierarchicalClustering1}, \cite{HierarchicalClustering2}. Unlike supervised learning and reinforcement learning that require training data and reward information exploration, respectively, to learn a particular model \cite{Reinforcement, Supervised}, unsupervised learning does not need such data or information, which is easy to be executed. 
Clustering methods have been proven to be able to improve the performance of spectrum sensing and wireless sensor networks \cite{Clustering1}, \cite{Clustering2}. According to our knowledge, this is the first work using clustering in data exchange. Instead of retrieving lost packets by retransmissions of macro BSs, a UAV will request its lost packets to other UAVs in the same cluster, so that lost packets retrieval could be conducted in a parallel fashion to enhance efficiency. By utilizing the agglomerative hierarchical clustering, appropriate UAVs will be clustered together, facilitating that UAVs in the same cluster are able to supply and supplement each other's lost packets. To further improve the efficiency of data exchange, a data exchange mechanism is developed to determine the order of UAVs in requesting their lost packets and replying received requests. The priorities of UAVs base on the number of its lost packets or the number of requested packets that it can provide in order to make each request-reply process provide maximum benefits to other UAVs' lost packets retrieval in the same cluster. In addition, a new structure of contention window is designed based on the carrier sense multiple access with collision avoidance (CSMA/CA) to support the data exchange mechanism.

The remainder of the paper is organized as follows. In Section II, system model considered in the paper is presented along with problem analysis of data exchange in UAV networks. The proposed data exchange scheme is introduced in Section III and Section IV, which elaborate the developed clustering method using unsupervised learning and data exchange mechanism, respectively. Section V showS simulation studies to demonstrate the effectiveness of the proposed scheme. Finally, Section VI concludes the whole paper.

\section{System Model and Problem Analysis}

In this paper, a UAV network, consisting of multiple UAVs, is considered, which serves as a small aerial BS. All UAVs in this network need to receive packets from a macro BS and these packets are commonly required for each UAV. Due to long-distance propagation and bad wireless environments, UAVs may encounter packet losses and their lost packets are retrieved by data exchange between UAVs rather than retransmissions of the BS. Considering the fact that UAVs are generally simple devices with limited power supply and control capabilities, the carrier-sense multiple access with collision avoidance (CSMA/CA) protocol is adopted to support spectrum access and communications between UAVs, which has low complexity and does not require centralized control \cite{802.11, 802.11n}.

With data exchange used to conduct the lost packet retrieval, two main and crucial technical issues are aroused. First, if multiple UAVs experience packet loss and need to request lost packets from other UAVs, which UAVs request their lost packets first is critical for the efficiency of the lost packet retrieval. Thus, the order of UAVs requesting their lost packets should be properly determined. Second, if multiple UAVs receive a request and hold at least one requested packets, which UAV is selected to reply this request should be studied.

For the first issue, instinctively a UAV with a large amount of lost packets should be allowed to request their lost packets first, as the corresponding reply comprised of requested lost packets is more likely to also benefit other UAVs and supply other UAVs' lost packets. On the other hand, the UAV, which owns the largest number of requested packets and can supply the requested packets as many as possible, should reply to the request, so that the requesting UAV obtains its lost packet with the minimum times of requests, meanwhile such a reply can also benefit other UAVs that may also request the same packets. To sum up, in our designed data exchange scheme in UAV networks, there are two basic rules, including:

\noindent $\bullet$ The UAV with a large amount of lost packets requests their lost packets first;

\noindent $\bullet$ After receiving a request, the UAV that possesses and can supply the maximum requested packets replies this request.

Fig. 1 gives a simple example, where it is assumed that there are four UAVs and six packets that need to be received by each UAV. Let $\mathbf{\Omega}  = \left\{ {{w_1},{w_2},{w_3},{w_4},{w_5},{w_6}} \right\}$ be the set of packets. As shown in Fig. 1, different UAVs received different subsets of $\mathbf{\Omega}$ and the lost packet retrieval is conducted by data exchange between UAVs. Based on the aforementioned analysis, it is better to let UAV 3 with the largest number of lost packets request its lost packets firstly. The requested lost packets of UAV 3 include $w_1$, $w_2$, $w_4$, and $w_6$. Apparently, UAV 4 can supply the largest number of requested packets for UAV 3. Then, UAV 4 will reply the request of UAV 3 with a reply packet carrying $w_1$, $w_2$, $w_4$, and $w_6$. It is noticeable that the reply of UAV 4 can also benefit UAV 1 and UAV 2, providing them with their lost packets, $w_6$ and $w_1$, respectively. After UAV 3's request and UAV 4's reply, UAV 2 and UAV 3 will receive the full set of transmitted packets. Afterwards, UAV 1 is the one with the maximum lost packets, which starts its request. Obviously, UAV 2 and UAV 3 with all the transmitted packets will be chosen to reply UAV 1's request, which can provide all the lost packets of UAV 1.

\vspace{-0em}\begin{figure}[t]
  \begin{center}
    \scalebox{0.75}[0.75]{\includegraphics{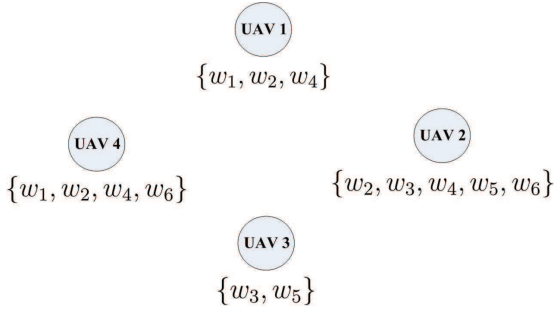}}
     \vspace{-0em}\caption{An example for problem analysis of data exchange in a UAV network.}
    \label{fig:1}
  \end{center}
\vspace{-0em}\end{figure}

With these rules, new technical issues emerge, which are: i) if numerous UAVs exist in a UAV network, a UAV may wait for a long time before it can request its lost packets, causing a long data exchange latency; ii) Based on CSMA/CA protocol, how to control the order of UAVs' request and reply according to the number of lost packets and owned requested packets, respectively, is required to be solved; iii) If multiple UAVs have the same number of lost packets or owned requested packets, there may be UAVs that request or reply simultaneously, incurring collisions with CSMA/CA protocol.

In the following two sections, a data exchange scheme based on clustering and unsupervised learning is developed to solve the remaining technical issues and realize efficient data exchange between UAVs. The developed data exchange scheme consists of two main components, clustering and data exchange in a cluster.

\section{Clustering with unsupervised learning}

Herein, the agglomerative hierarchical clustering, an unsupervised learning method, is employed to partition a UAV network to multiple clusters. A UAV can only be included in one cluster and perform data exchange with UAVs in the same cluster. To avoid interference and spectrum competitions between clusters, frequency division is adopted that UAVs in different clusters use different wireless spectrum to communicate. In the tradition agglomerative hierarchical clustering, each object (UAV in this paper) is deemed as a cluster at the beginning. Then, calculate the similarity or dissimilarity of every pair of clusters, which is characterized by the distance between clusters. The two clusters with the smallest or largest distance are merged to a cluster. Repeat the merging process until the number of cluster reaches a desired value \cite{HierarchicalClustering1, HierarchicalClustering2}. Apparently, a significant defect of the tradition agglomerative hierarchical clustering is the low flexibility, where the number of clusters may decrease exponentially. This may result in the number of clusters far away from the desired number or the imbalance of number of UAVs in different clusters.

To cope with that, a novel agglomerative clustering method is designed for the clustering of UAVs. First, the distance measure between two objects needs to be defined. Let $M$ be the total number of packets that are required to be received by all UAVs. A packet indicator vector is defined according to the packets that an object has already obtained. An object could be a UAV or a cluster. If an object is a cluster, its packet indicator vector should reflect the packets that are collected by all the UAVs in the cluster. The packet indicator vector of UAV $i$ can be expressed as ${\mathbf{v}_i} = \left\{ {{\rho _{i,1}},{\rho _{i,2}}, \cdot  \cdot  \cdot  \cdot  \cdot  \cdot ,{\rho _{i,M}}} \right\}$, where ${\rho _{i,m}} = 1$ if packet $m$ has already been attained by UAV $i$, otherwise ${\rho _{i,m}} = 0$, $m = 0,1, \cdot  \cdot  \cdot  \cdot  \cdot  \cdot ,M$. On the other hand, for a cluster $n$, its packet indicator vector is in the form same with that of a UAV, ${\mathbf{v}_n} = \left\{ {{\rho _{n,1}},{\rho _{n,2}}, \cdot  \cdot  \cdot  \cdot  \cdot  \cdot ,{\rho _{n,M}}} \right\}$, while the definition of ${\rho _{n,m}}$ is different, where ${\rho _{n,m}}=1$ if at least one UAV in cluster $n$ possess packet $m$, otherwise ${\rho _{n,m}}=0$. The distance between two objects, $i$ and  $j$, is represented by $dist\left( {i,j} \right)$. In this paper, hamming distance is adopted to calculate distances between two objects, which is the number of data positions in two binary data vectors, where two corresponding data are different \cite{Hamming}. It is obvious that $dist\left( {i,j} \right)$ would be larger if $i$ and $j$ have a bigger difference in their owned packets. In other words, a larger $dist\left( {i,j} \right)$ can be acquired if $i$ and $j$ hold and can supply more each other's lost packets.

With the defined packet indicator vector and distance, the process of UAV clustering is designed, which include two stages, an initialization stage and a merging stage. It is assumed that the desired number of clusters is $N$. Undoubtedly, a small distance between two UAVs indicates that these two UAVs have similar received packets and can rarely contribute to other UAV's lost packet retrieval, so that UAVs with small distances should be assigned to different clusters. In the initialization stage, $N$ UAVs will be selected out, each of which will be deemed as an individual cluster. The process of the initialization stage is described as follows. Find a UAV-UAV pair with a minimum distance from the UAV pool. Then, two UAVs in this UAV-UAV pair are included in two different clusters, which will also be removed from the UAV pool. Afterwards, find another UAV-UAV pair with a minimum distance from the UAV pool and the corresponding UAVs are assigned to two new different clusters. Repeat this process until $N$ UAVs are found out and form $N$ clusters. Note that in the initialization stage clusters are initialized by selecting out appropriate UAV-UAV pairs. As a result, the number of found UAVs is always even, while $N$ may be odd. In the case of an odd $N$, $N+1$ UAVs will be selected out with the aforementioned process. To initialize $N$ clusters, only randomly chosen $N$ UAVs from those $N+1$ UAVs are remained to be $N$ different clusters, while the rest one will be back to the UAV pool.

The merging stage proceeds by conducting merging iterations until all UAVs have been assigned to clusters. In a merging iteration, the detailed process is elaborated as follows with the notations of $\mathbf{\Phi}$, $\mathbf{\Delta}$, and $\mathbf{C}_n$, $n = 1,2, \cdot  \cdot  \cdot ,N$, representing the set of clusters, the UAV pool comprising UAVs that have not been assigned to any clusters, and the set of UAVs assigned to cluster $n$, respectively. First, $\mathbf{\Phi}$ is set to be $\mathbf{\Phi}  = \left\{ {1,2, \cdot  \cdot  \cdot  \cdot  \cdot  \cdot ,N} \right\}$ and the distances between clusters and the UAVs that are still in $\mathbf{\Delta}$, are calculated. According to calculated distances, find a cluster-UAV pair $\left( {{n^*},{i^*}} \right)$ with the largest distance, ${n^*},{i^*} = \arg \,\mathop {\max }\limits_{n \in \mathbf{\Phi} ,i \in \mathbf{\Delta} } dist\left( {n,i} \right)$. Then, remove ${n^*}$ and $i^*$ from $\mathbf{\Phi}$ and $\mathbf{\Delta}$ by $\mathbf{\Phi}  = \mathbf{\Phi}  - {n^*}$ and $\mathbf{\Delta}  = \mathbf{\Delta}  - {i^*}$, respectively, and $i^*$ is included in the cluster $n^*$ by ${\mathbf{C}_{{n^*}}} = {\mathbf{C}_{{n^*}}} \cup \left\{ {{i^*}} \right\}$. The current merging iteration will complete until $\mathbf{\Phi}  = \emptyset$ or $\mathbf{\Delta}  = \emptyset$, indicating that each cluster has already been assigned with a UAV in the current merging iteration or there is no UAV that has not been assigned. Based on the definition of the distance, a large distance reflects the dissimilarity of owned packets. Merging a cluster with a UAV in a large distance can facilitate that UAVs, which are able to supply and supplement each other's lost packets, are grouped into the same cluster. After finishing the current merging iteration, all the clusters' packet indicator vectors are updated according to their new included UAVs by conducting or operation for packet indicator vectors of a cluster and its new UAV members. For example, assume that the current packet indicator vector of cluster $n$ is ${\mathbf{v}_n} = \left\{ {1,0,0,1,0,0} \right\}$ and UAV $i$ with the packet indicator vector ${\mathbf{v}_i} = \left\{ {0,1,0,0,1,0} \right\}$ is assigned to $n$. Accordingly, the updated packet indicator vector of cluster $n$ should be ${\mathbf{v}_n} = \left\{ {1,1,0,1,1,0} \right\}$. The next merging iteration starts with resetting $\mathbf{\Phi}$ to be $\mathbf{\Phi}  = \left\{ {1,2, \cdot  \cdot  \cdot  \cdot  \cdot  \cdot ,N} \right\}$ and updating the distances between clusters and the UAVs in $\mathbf{\Delta}$. Merging iterations are repeatedly carried out until there is no UAV in the UAV pool and all UAVs have been included in clusters, namely $\mathbf{\Delta}  = \emptyset$.

Obviously, instead of merging two clusters as one in the traditional agglomerative clustering, our proposed clustering method merges a cluster with only one UAV in once merging. As a result, the merging process is carried out gradually and keeps clusters in a desired amount. Moreover, by partitioning the whole UAV network into multiple clusters, the limited amount of UAVs will be included in a cluster, conducting data exchange within the cluster, which are able to provide and complement each other's lost packets. By this way, the waiting time of UAVs for requesting their lost packets and the data exchange delay can be effectively reduced.

In summary, the overall clustering process with the proposed agglomerative clustering method is given in the algorithm 1.

\begin{algorithm}
\caption{The clustering process using the designed agglomerative clustering.}
\label{alg1}
\begin{algorithmic}
\STATE {\textbf{1. Clustering Initialization:}\\
1.1 Include all UAVs in $\mathbf{\Delta}$. Calculate distances between UAVs with the hamming distance. Set ${\mathbf{C}_n} = \emptyset$, $n = 1,2, \cdot  \cdot  \cdot ,N$.\\
1.2 Find a UAV-UAV pair with a minimum distance.\\
1.3 Remove two UAVs of the UAV-UAV pair from $\mathbf{\Delta}$.\\
1.4 Repeat 1.2 and 1.3 until $N$ UAVs are found out.\\
1.5 Each of found $N$ UAVs is deemed as an individual cluster.\\
1.6 Update $\mathbf{\Delta}$ and $\mathbf{C}_n$.
}

\STATE {\textbf{2. Merging iterations:}\\
2.1 Let $\mathbf{\Phi}$ be $\mathbf{\Phi}  = \left\{ {1,2, \cdot  \cdot  \cdot  \cdot  \cdot  \cdot ,N} \right\}$. \\

2.2 Compute the distances between clusters and UAVs still in $\mathbf{\Delta}$.\\
2.3 Find the cluster-UAV pair $\left( {{n^*},{i^*}} \right)$ with the largest distance, ${n^*},{i^*} = \arg \,\mathop {\max }\limits_{n \in \mathbf{\Phi} ,i \in \mathbf{\Delta} } dist\left( {n,i} \right)$.\\
2.4 Update $\mathbf{\Phi}$, $\mathbf{\Delta}$, and $\mathbf{C}_n$ by $\mathbf{\Phi}  = \mathbf{\Phi}  - {n^*}$, $\mathbf{\Delta}  = \mathbf{\Delta}  - {i^*}$, and ${\mathbf{C}_{{n^*}}} = {\mathbf{C}_{{n^*}}} \cup \left\{ {{i^*}} \right\}$, respectively.\\
2.5 Repeat 2.2, 2.3, and 2.4 until $\mathbf{\Phi}  = \emptyset$ or $\mathbf{\Delta}  = \emptyset$.\\
2.6 Update packet indicator vectors of all clusters, $\mathbf{v}_n$, $n = 1,2, \cdot  \cdot  \cdot ,N$.
}

\STATE{\textbf{3. Clustering results:}
Repeat merging iterations until $\mathbf{\Delta}  = \emptyset $. Then, output the clustering results, ${\mathbf{C}_n}$, $n = 1,2, \cdot  \cdot  \cdot ,N$.
 }
\end{algorithmic}
\end{algorithm}

\section{Data exchange mechanism in a cluster}

By designing data exchange mechanism for UAVs in a cluster, three purposes intend to be achieved, including i) the UAV with the largest amount of lost packets has priority to request their lost packets; ii) the UAV that possesses the maximum requested packets replies the request first; iii) avoid spectrum access collision when UAVs have the same priority to request or reply. Due to the simple device nature, in this paper it is assumed that UAVs are configured with the CSMA/CA protocol to support wireless transmissions between UAVs. With that, UAVs compete for spectrum access through the random backoff procedure. Before a UAV attempts to access spectrum and send data, it chooses a random backoff time and starts countdown. During the backoff time, the UAV continues to sense wireless spectrum. If spectrum is detected to be busy, the backoff time count will be suspended. The backoff time count is resumed when spectrum becomes idle again. The UAV access the spectrum only if the backoff time counts to 0 \cite{802.11, 802.11n}.

Inspired by the fact that Wi-Fi systems configure data frames, control frames, and management frames with different spectrum access priorities by adjusting their backoff time, we design a data exchange mechanism based on a new random backoff procedure, where the backoff time of a UAV is inversely proportional with the current amount of lost packets that have failed to receive and not been retrieved through data exchange or the number of requested packets that a potential replier possesses. Recall that $M$ is the number of packets required to be received by all UAVs. As shown in Fig. 2, the length of the maximum contention window, the maximum backoff time, is $T$, which is equally divided into $M$ subwindows. The time length of each subwindow is $\frac{T}{M}$ and the time range of subwindow $K$, $k = 1,2, \cdot  \cdot  \cdot ,M$, is $\left( {\left. {\frac{{\left( {k - 1} \right)T}}{M},\frac{{kT}}{M}} \right]} \right.$. With the proposed random backoff procedure, if a UAV currently has $M'$ lost packets or possesses $M'$ requested packets, $M' \leqslant M$, its backoff time will be generated by randomly selecting a value within subwindow $M - M' + 1$. It is apparent that this random backoff procedure can offer a higher priority for UAVs with more lost packets or more requested packets to access spectrum. Additionally, UAVs with the same priority also have different backoff time to avoid spectrum access collision, as their backoff times are randomly selected within the same subwindow.

\vspace{-0em}\begin{figure}[t]
  \begin{center}
    \scalebox{0.75}[0.75]{\includegraphics{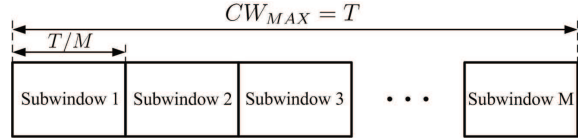}}
     \vspace{-0em}\caption{The designed random backoff procedure in data exchange.}
    \label{fig:1}
  \end{center}
\vspace{-0em}\end{figure}

Based on the random backoff procedure, the data exchange mechanism is designed as follows. For a UAV that needs to request its lost packets, it generates the backoff time according to its lost packet number and starts to count. If wireless spectrum becomes busy, it will suspend the backoff time count and continue to sense wireless spectrum until spectrum is sensed to be idle. During the suspension and spectrum sensing, it may receive a packet from a UAV in the same cluster. If the received packet is a reply packet, the UAV will read the reply packet and check if the reply packet carries its lost packets. If so, the UAV receives its lost packets from the received reply packet and updates its request information. On the other hand, if the received packet is a request packet, the UAV's suspension and spectrum sensing will continue until a reply packet responding to the request packet is transmitted. According to the reply packet, the UAV receives its required packets and updates its request information. Once the backoff time is finally counted to 0, the UAV will access spectrum and send its request. For UAVs replying a request, they attain their backoff time according to the requested packets that they possessed, and then start to count. During the backoff, once they found the request has been replied by another UAV, they will not reply the request anymore. Clearly, in a cluster, the request and reply can be conducted in a successive fashion. 
This would enable UAVs to take fully advantage of each request and reply to retrieve their lost packets, so that a lost packet will not be repetitively requested and replied in a cluster for fast lost packet retrieval and low overhead.

\section{Simulation Results}

In this section, extensive simulations are conducted to show the performance of the developed unsupervised learning-enabled data exchange scheme in terms of the number of data exchange and data exchange delay. Clearly, the number of data exchange is an important performance indicator directly determining data time consumption and overhead of data exchange, while the simulation of data exchange delay is used to testify if the proposed data exchange scheme can provide efficient data exchange. In this paper, we consider an practical scenario that no powerful infrastructure exists to support centralized control for wireless communications between UAVs. CSMA/CA is able to handle wireless transmissions between nodes in a distributed fashion without centralized control and has been widely applied in Wi-Fi systems. Thus, data exchange using CSMA/CA will be adopted as benchmark for performance comparison.

Additionally, simulations are conducted under two assumptions for simplification. First, in the application of UAV networks working as a small aerial BS, UAVs gather together working as a swarm to handle complicated missions. Thus, it is assumed that all UAVs in a UAV network share same channel states between UAVs and a macro BS, which UAVs periodically receive packets from. Correspondingly, wireless transmissions between the macro BS and all UAVs share the same packet delivery rate, $\rho$, where the packet delivery rate is defined as the probability that a UAV could successfully receive a packet from the BS. Second, packet loss of wireless transmissions between UAVs is neglected due to short-distance transmissions. Note that simulations of all the methods considered in performance comparison are conducted under the same assumptions. Therefore, these two assumptions will not have an negative effect on demonstrating the effectiveness and superiority of our proposed scheme.

\subsection{Optimal number of clusters}

First, the optimal number of clusters should be investigated for the unsupervised learning-based clustering of our developed scheme. Obviously, a large number of clusters could result in a more efficient data exchange between UAVs, the process of which will simultaneously be carried out in different clusters. However, more clusters will also cause that less UAVs included in a cluster, bringing in a risk that UAVs in a cluster do not collect a full set of packets. In other words, a UAV may not supplement its lost packets just through data exchange within its cluster. Hence, given a set of system parameters, including the number of UAVs, the number of packets in data exchange, and a packet delivery rate, $\rho$, the optimal number of clusters will be found out through simulation studies.

\vspace{-0em}\begin{figure}[t] 
  \begin{center}
    \scalebox{0.6}[0.6]{\includegraphics{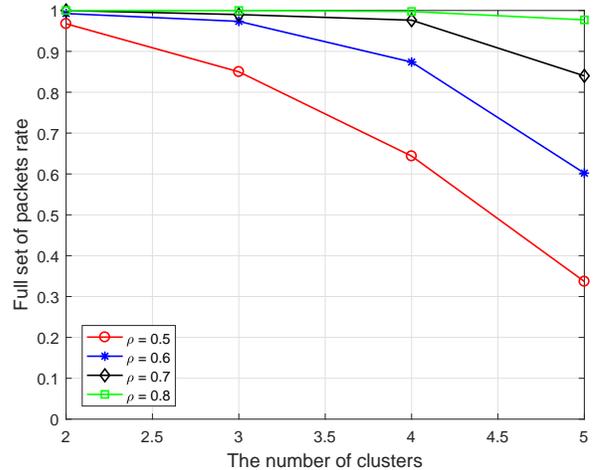}}
     \vspace{-0em}\caption{Full set of packets rate versus the number of clusters in a UAV network with 10 UAVs and 6 packets.}
    \label{fig:1}
  \end{center}
\vspace{-0em}\end{figure}

\vspace{-0em}\begin{figure}[t] 
  \begin{center}
    \scalebox{0.6}[0.6]{\includegraphics{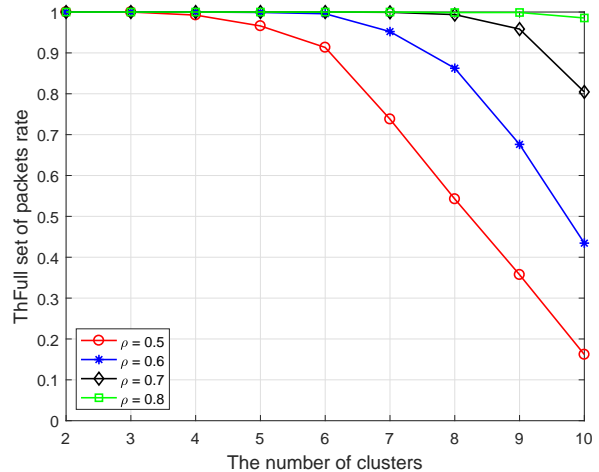}}
     \vspace{-0em}\caption{Full set of packets rate versus the number of clusters in a UAV network with 20 UAVs and 10 packets.}
    \label{fig:1}
  \end{center}
\vspace{-0em}\end{figure}

Fig. 3 shows the full set of packets rate in a network with 10 UAVs and 6 packets under varying packet delivery rates, $\rho$, which is a ratio of clusters owning a full set of packets to all clusters. For each set of system parameters, 500 simulations are conducted to show universal and overall performance. The general trends of Fig. 3 is that the full set of packets rate declines with the increase of clusters and the decrease of $\rho$. This is because a larger number of clusters makes less UAVs assigned to a cluster and a smaller $\rho$ reduces packets possessed by each individual UAV, both of which will compromise the chance of a cluster collecting a full set of packets by UAV clustering. Fig. 4 plots the full set of packets rate in a network with 20 UAVs and 10 packets. Apparently, it has the similar trends with Fig. 3.

The design insights of our proposed scheme could be given with Fig. 3 and Fig. 4. Given a set of system parameters, the maximum number of clusters with a very high full set of packets rate should be selected as the optimal number of clusters. This number of clusters can make as many as UAVs retrieving their lost packets simultaneously with a high data exchange efficiency, while guaranteeing that all lost packets be can supplied by data exchange within a cluster. For example, for a network with 10 UAVs, 6 packets, and $\rho = 0.7$, the optimal number of clusters should be 3, where the full set of packets rate is near to 1. For a network with 20 UAVs, 10 packets, and $\rho = 0.6$, the optimal number of clusters is 6. Additionally, if a UAV network is in a bad wireless environment with a small $\rho$, for example when $\rho = 0.5$ in Fig. 3, clustering method is inapplicable, as even only partitioning a network into 2 clusters cannot ensure that a cluster collect a full set of packets.

\subsection{Data exchange number}

Here, simulation results of data exchange numbers that are conducted to make all UAVs obtain a full set of packets are presented. Once data exchange is defined as a request-reply process. Fig. 5 and Fig. 6 show the data exchange number in a UAV network with 10 UAVs, 6 packets, and $\rho = 0.7$, and in another UAV network with 20 UAVs, 10 packets, and $\rho = 0.6$, respectively. According to Fig. 3 and Fig. 4, in such two networks, their optimal cluster numbers are 3 and 6, respectively. Thus, in the simulations of the proposed data exchange scheme, unsupervised learning-based clustering will respectively result in 3 and 6 clusters. Simulation results shown in Fig. 5 and Fig. 6 are average values of 500 simulations. Note that as with the proposed scheme multiple clusters perform data exchange in a parallel fashion, simulation results of the proposed scheme are data exchange numbers of the cluster with the maximum data exchanges, which determines the completion and delay of the whole data exchange process. From Fig. 5 and Fig. 6, it can be observed that the proposed scheme with both our designed clustering method and data exchange mechanism has the lowest data exchange number owing to the fact that the use of clustering could allow multiple UAVs to carry out data exchange simultaneously. In addition, it is obvious that even only using the designed data exchange mechanism without clustering can significantly improve the efficiency of data exchange, as UAVs will exchange packets in a more appropriate order, making each request-reply process not only satisfy the request but also provide maximum benefits to other UAVs. By comparing Fig. 5 and Fig. 6, it shows that in a worse wireless environment with a smaller $\rho$, more data exchange have to be experienced with CSMA/CA, while utilizing proper clustering and data exchange mechanism can effectively alleviate the negative impact brought by a smaller $\rho$ and make the whole data exchange process completed with limited request-reply processes.

\vspace{-0em}\begin{figure}[t] 
  \begin{center}
    \scalebox{0.6}[0.6]{\includegraphics{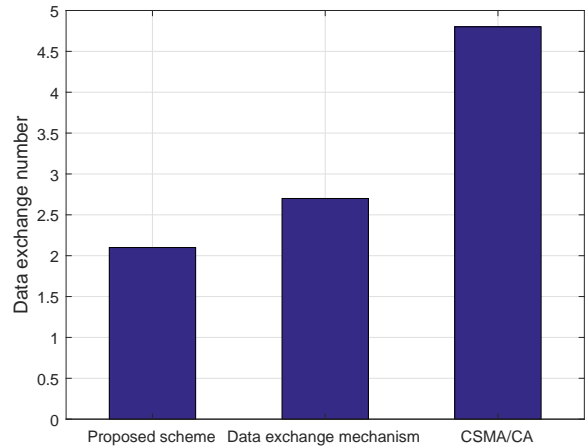}}
     \vspace{-0em}\caption{Data exchange number with 10 UAVs, 6 packets, $\rho = 0.7$, and 3 clusters.}
    \label{fig:1}
  \end{center}
\vspace{-0em}\end{figure}

\vspace{-0em}\begin{figure}[t] 
  \begin{center}
    \scalebox{0.6}[0.6]{\includegraphics{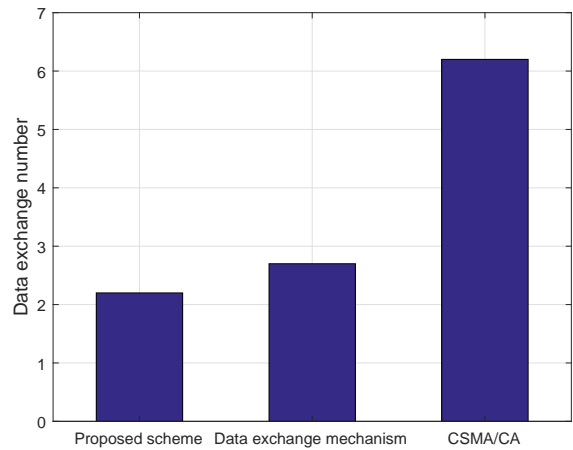}}
     \vspace{-0em}\caption{Data exchange number with 20 UAVs, 10 packets, $\rho = 0.6$, and 6 clusters.}
    \label{fig:1}
  \end{center}
\vspace{-0em}\end{figure}

\subsection{Data exchange delay}

Data exchange delay mainly is caused by spectrum access contentions and packet transmissions. According to the specifications of 802.11n \cite{802.11n}, the sensing time of 802.11n is from DCF (distributed coordination function) inter-frame space (DIFS) to DIFS plus maximum contention window length, $CW_{MAX}$, namely from 34 $\mu s$ to 9241 $\mu s$. Accordingly, in the simulation, the $CW_{MAX}$ is set to be 9241 $\mu s$. The spectrum access contention of our proposed scheme is calculated based on the random backoff procedure designed in Fig. 2. The delay of packet transmissions is proportional to the number of packets included in the data field of a 802.11 physical packet structure. For simplification, it is assumed that each packet is 2 $ms$. The packet transmission delay is $2K$ $ms$, if $K$ packets are included in a packet. Additionally, in each 802.11 physical packet, the preamble, as packet header, also bring in additional delay, normally including a $8$ $\mu s$ short training field, a $8$ $\mu s$ long training field, and a $4$ $\mu s$ signal field.

Fig. 7 and Fig. 8 plot the data exchange delay under the same system parameters with Fig. 5 and Fig. 6, respectively. All the simulation results are average values of 500 simulations. By taking fully each request-reply opportunity to supply lost packets not only to the request UAV but also to other UAVs in the same cluster, the designed data exchange mechanism can effectively reduce data exchange delay. Proper clustering using unsupervised learning can further improve the performance on data exchange delay owning to parallel packet exchanges in multiple clusters.

\vspace{-0em}\begin{figure}[t] 
  \begin{center}
    \scalebox{0.6}[0.6]{\includegraphics{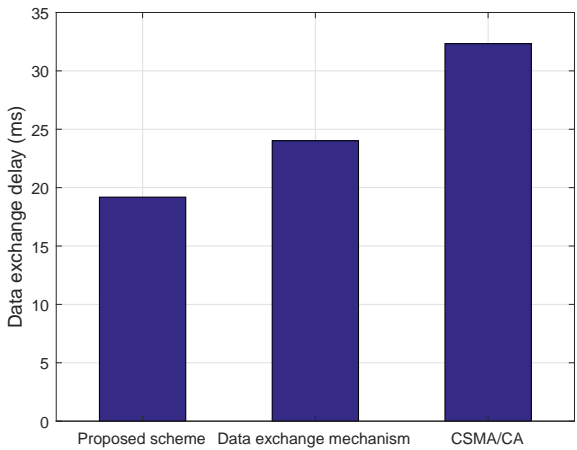}}
     \vspace{-0em}\caption{Data exchange delay with 10 UAVs, 6 packets, $\rho = 0.7$, and 3 clusters.}
    \label{fig:1}
  \end{center}
\vspace{-0em}\end{figure}

\vspace{-0em}\begin{figure}[t] 
  \begin{center}
    \scalebox{0.6}[0.6]{\includegraphics{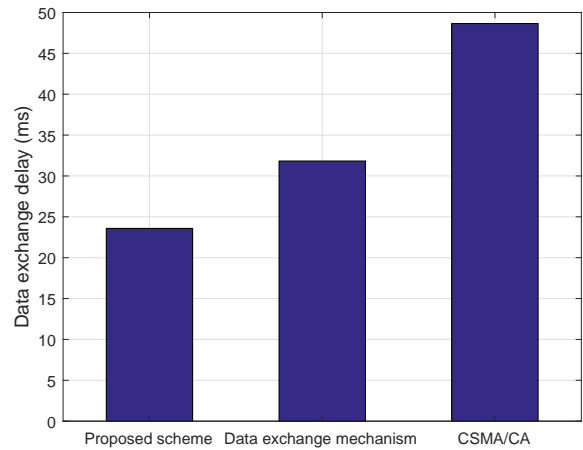}}
     \vspace{-0em}\caption{Data exchange delay with 20 UAVs, 10 packets, $\rho = 0.6$, and 6 clusters.}
    \label{fig:1}
  \end{center}
\vspace{-0em}\end{figure}

\section{Conclusions}

Compared to retrieving lost packets through retransmissions of macro BSs, lost packet retrieval through data exchange would be more efficient with less overhead and energy consumption. In this paper, a novel data exchange scheme is developed with two main components, a clustering approach and a data exchange mechanism. The agglomerative hierarchical clustering, a type of unsupervised learning, is employed to design the clustering approach, where UAVs are assigned to different clusters according to their received packets and UAVs that can supply each other's lost packets will be clustered together. In the designed data exchange mechanism, unlike UAVs carry out data exchange in a random order with tradition methods, like CSMA/CA, the order of UAVs executing data change is optimized according to the number of their lost packets or requested packets that they can provide. By this way, each request-reply process can not only maximize the lost packet retrieval of the UAV sending request, but also help other UAVs in the same cluster retrieve their lost packets. To avoid the undesired situation that UAVs in a cluster do not possess a full set of packets, the optimal number of clusters is studied for UAVs networks with different system parameters. Then, under the optimal number of clusters, simulation studies indicate that our developed data exchange scheme can significantly improve the performance on data exchange numbers and delay.



\begin{thebibliography}{99}

\bibitem{UAVSurvery1}
Y. Zeng, R. Zhang, and T. J. Lim, ``Wireless Communications with Unmanned Aerial Vehicles: Opportunities and Challenges," \emph{IEEE Communications Magazine}, vol. 54, no. 5, pp. 36-42, May 2016.

\bibitem{UAVSurvery2}
M. Mozaffari, W. Saad, M. Bennis, Y. Nam, and M. Debbah, ``A Tutorial on UAVs for Wireless Networks: Applications, Challenges, and Open Problems," \emph{IEEE Communications Surveys \& Tutorials}, vol. 21, no. 3, pp. 2334-2360, Mar 2019.

\bibitem{UAVSurvery3}
B. Shang, L. Liu, J. Ma, and P. Fan, ``Unmanned Aerial Vehicle Meets Vehicle-to-Everything in Secure Communications," \emph{IEEE Communications Magazine}, vol. 57, no. 10, pp. 98-103, Oct 2019.


\bibitem{UAVEmergency}
N. Zhao, W. Lu, M. Sheng, Y. Chen, J. Tang, F. Yu, and K. K. Wong, ``UAV-Assisted Emergency Networks in Disasters," \emph{IEEE Wireless Communications}, vol. 26, no. 1, pp. 45-51, Feb 2019.

\bibitem{UAVEmergency1}
K. G. Panda, S. Das, D. Sen, and W. Arif, ``Design and Deployment of UAV-aided Post-disaster Emergency Network," \emph{IEEE Access}, vol. 7, pp. 102985¨C102999, 2019.

\bibitem{UAVSensor}
J. Baek, S. I. Han, and Y. Han, ``Energy-Efficient UAV Routing for Wireless Sensor Networks," \emph{IEEE Transactions on Vehicular Technology}, vol. 69, no. 2, pp. 1741-1750, Feb 2020.

\bibitem{UAVRoutineMission1}
J. Park, S. Choi, I. Ahn, and J. Kim, ``Multiple UAVs-based Surveillance and Reconnaissance System Utilizing IoT Platform," \emph{2019 International Conference on Electronics, Information, and Communication (ICEIC)}, Auckland, New Zealand, pp. 1-3, 2019.

\bibitem{UAVRoutineMission2}
P. Nintanavongsa and I. Pitimon, ``Impact of Sensor Mobility on UAV-based Smart Farm Communications," \emph{2017 International Electrical Engineering Congress (iEECON)}, Pattaya, pp. 1-4, 2017.


\bibitem{UAVCellular}
S. Sekander, H. Tabassum, and E. Hossain, ``Multi-Tier Drone Architecture for 5G/B5G Cellular Networks: Challenges, Trends, and Prospects," \emph{IEEE Communications Magazine}, vol. 56, no. 3, pp. 96-103, Mar 2018.


\bibitem{Retransmission1}
H. Ding, S. Ma, C. Xing, Z. Fei, Y. Zhou, and C. L. P. Chen, ``Analysis of Hybrid ARQ in Ad Hoc Networks with Correlated Interference and Feedback Errors," \emph{IEEE Transactions on Wireless Communications}, vol. 12, no. 8, pp. 3942-3955, Aug 2013.

\bibitem{Retransmission2}
K. Chi, Z. Yu, Y. Li, and Y. Zhu, ``Energy-Efficient D2D Communication Based Retransmission Scheme for Reliable Multicast in Wireless Cellular Network," \emph{IEEE Access}, vol. 6, pp. 31469-31480, 2018.

\bibitem{Retransmission3}
H. Ding, Z. Shi, S. Ma, and C. Xing, ``On the Performance of HARQ-IR Over Nakagami-m Fading Channels in Mobile Ad Hoc Networks," \emph{IEEE Transactions on Vehicular Technology}, vol. 66, no. 5, pp. 3913-3929, May 2017.

\bibitem{SwarmUAV1}
H. Song, L. Liu, S. Pudlewski, and E. S. Bentley, ``Random Network Coding Enabled Routing Protocol in Unmanned Aerial Vehicle Networks," \emph{IEEE Transactions on Wireless Communications}, vol. 19, no. 12, pp. 8382-8395, Dec 2020.

\bibitem{SwarmUAV2}
H. Song, L. Liu, S. Pudlewski, and E. S. Bentley, ``Random Network Coding Enabled Routing in Swarm Unmanned Aerial Vehicle Networks," \emph{2019 IEEE Global Communications Conference (GLOBECOM)}, Waikoloa, HI, USA, pp. 1-6, 2019.

\bibitem{SwarmUAV3}
S. Hayat, E. Yanmaz and R. Muzaffar, ``Survey on Unmanned Aerial Vehicle Networks for Civil Applications: A Communications Viewpoint," \emph{IEEE Communications Surveys \& Tutorials}, vol. 18, no. 4, pp. 2624-2661, 2016.


\bibitem{HierarchicalClustering1}
A. Bouguettaya, Q. Yu, X. Liu, X. Zhou, and A. Song, ``Efficient agglomerative hierarchical clustering," \emph{Expert Systems with Applications}, vol. 42, no. 5, pp. 2785-2797, 2015.

\bibitem{HierarchicalClustering2}
M. Roux, ``A Comparative Study of Divisive and Agglomerative Hierarchical Clustering Algorithms," \emph{Journal of Classification} vol.35, no. 2, pp. 345-366, 2018.

\bibitem{Reinforcement}
H. Song, J. Bai, Y. Yi, J. Wu, and L. Liu, ``Artificial Intelligence Enabled Internet of Things: Network Architecture and Spectrum Access," \emph{IEEE Computational Intelligence Magazine}, vol. 15, no. 1, pp. 44-51, Feb 2020.

\bibitem{Supervised}
Z. Zhou, L. Liu, S. Jere, J. Zhang and Y. Yi, ``RCNet: Incorporating Structural Information into Deep RNN for Online MIMO-OFDM Symbol Detection with Limited Training," \emph{IEEE Transactions on Wireless Communications}, accepted for publication, 2021. 


\bibitem{Clustering1}
B. Shang, L. Liu, H. Chen, J. Zhang, S. Pudlewski, E. S. Bentley, and J. D. Ashdown, ``Spatial Spectrum Sensing in Uplink Two-Tier User-Centric Deployed HetNets," \emph{IEEE Transactions on Wireless Communications}, vol. 19, no. 12, pp. 7957-7972, Dec 2020.

\bibitem{Clustering2}
V. Pal, G. Singh, and R. P. Yadav, ``Balanced Cluster Size Solution to Extend Lifetime of Wireless Sensor Networks," \emph{IEEE Internet of Things Journal}, vol. 2, no. 5, pp. 399-401, Oct 2015.


\bibitem{802.11n}
E. Perahia and R. Stacey, ``Next Generation Wireless LANs: Throughput, Robustness, and Reliability in 802.11n," \emph{Cambridge University Press}, Dec 2009.

\bibitem{802.11}
M. Gast, ``802.11 Wireless Networks: The Definitive Guide," O'Reilly Media, Inc., 2005.

\bibitem{Hamming}
C. Yao, P. Chen, T. Wang, Y. S. Han, and P. K. Varshney, ``Performance Analysis and Code Design for Minimum Hamming Distance Fusion in Wireless Sensor Networks," \emph{IEEE Transactions on Information Theory}, vol. 53, no. 5, pp. 1716-1734, May 2007.

\end{thebibliography}
\end{document}